\documentclass[12pt, draftclsnofoot, onecolumn]{IEEEtran}
\usepackage{setspace}
\usepackage{cite}

\usepackage{graphicx}
\graphicspath{/figs/}
\DeclareGraphicsExtensions{.pdf,.jpeg,.png}
\usepackage{amsmath, amssymb}

\usepackage[table]{xcolor} 
\usepackage{bm}

\usepackage{float}
\usepackage[caption = false]{subfig}
\usepackage{algorithm}
\usepackage{algorithmicx}
\usepackage{algpseudocode}
\usepackage{pifont}
\usepackage{xcolor}
\usepackage{relsize}

\usepackage{array}
\usepackage{url}
\usepackage{parskip}
\usepackage{layout}

\let\OLDthebibliography\thebibliography
\renewcommand\thebibliography[1]{
  \OLDthebibliography{#1}
  \setlength{\parskip}{0pt}
  \setlength{\itemsep}{4dd}
}

\usepackage{mathtools}

\DeclareMathOperator*{\argmax}{arg\,max}

\algdef{SE}[DOWHILE]{Do}{doWhile}{\algorithmicdo}[1]{\algorithmicwhile\ #1}%

\newcommand{\ue}{u \color{black}}

\newcommand{\analog}[1]{#1^{\text{RF}}}
\newcommand{\Nxa}{\analog{N}_{\text{X}}}
\newcommand{\Nya}{\analog{N}_{\text{Y}}}

\newcommand{\Nx}{N_{\text{X}}}
\newcommand{\Ny}{N_{\text{Y}}}
\newcommand{\Nt}{N_{\text{T}}}
\newcommand{\Nta}{\analog{N}_{\text{T}}}
\newcommand{\Nr}{N_{\text{R}}}

\newcommand{\nt}{n_{\text{t}}}
\newcommand{\nr}{n_{\text{r}}}

\newcommand{\norm}[1]{\left\lVert#1\right\rVert}

\newcommand{\vect}[1]{\mathbf{#1}}
\newcommand{\complex}{\mathbb{C}}

\newcommand{\phaseres}{b_{\text{phase}}}

\newcommand{\F}{\mathbf{F}}

\newcommand{\f}{\mathbf{f}}

\newcommand{\Fa}{\analog{\mathbf{F}}_{c, t}}
\newcommand{\Fap}{\analog{\mathbf{F}}_{c', t}}

\newcommand{\Fistar}{\F_{b_u, \hat{i}_{b_u, u}}}
\newcommand{\Fssb}{\mathbf{F}^{\text{SSB}}}

\newcommand{\Bssb}{\mathbf{B}^{\text{SSB}}}
\newcommand{\Bcsi}{\mathbf{B}^{\text{CSI-RS}}}

\newcommand{\Bactive}{\mathbf{B}^{\text{sub}}}
\newcommand{\Str}{\vect{s}^{\text{tr}}_{c, t, k}}
\newcommand{\Lmax}{L_{\text{{max}}}}
\newcommand{\Kssb}{K^{\text{SSB}}}
\newcommand{\Tssb}{T^{\text{SSB}}_{i}}

\newcommand{\ycsirs}{\vect{y}^{\text{CSI-RS}_i}}
\newcommand{\Kcsi}{K^{\text{CSI-RS}_i}}
\newcommand{\Tcsi}{T^{\text{CSI-RS}_{i}}}
\newcommand{\Lcsi}{L_{\text{{CSI}}}}
\newcommand{\bwp}{\text{S}_{\text{B}}}

\newcommand{\Ncsi}{N_{\text{CSI}}}
\newcommand{\Bg}{B_{\text{g}}}
\newcommand{\ihat}{\widehat{\vect{i}}_{c, u}}

 \newcommand{\Obsc}{\vect{O}^{\text{BSC}}}
 \newcommand{\Nxo}{N_{\text{X,O}}}
  \newcommand{\Nyo}{N_{\text{Y,O}}}
  \newcommand{\Fssbpred}{\widehat{\mathbf{F}}^{\text{SSB}}}
\newcommand{\Fcsipred}{\widehat{\mathbf{F}}^{\text{CSI-RS}}}
\newcommand{\Achan}{\vect{H}}

\newcommand{\algo}{\text{NBL}}
\newcommand{\tku}[1]{{#1_{\ue, t, k}}}

\newcommand{\RSRPciu}{\text{RSRP}_{c, i, u}}

\newcommand{\SINR}{\text{SINR}}

\begin{document}

\title{Neural Codebook Design for Network Beam Management}

\author{\IEEEauthorblockN{Ryan M. Dreifuerst,~\IEEEmembership{Graduate Student Member,~IEEE,} and }%
    \and
	\IEEEauthorblockN{Robert~W.~Heath~Jr.~\IEEEmembership{Fellow,~IEEE}}%

	\thanks{Ryan M. Dreifuerst is with North Carolina State University, Raleigh, NC 27695 (rmdreifu@ncsu.edu). Robert W. Heath Jr. is with the University of California, San Diego (rwheathjr@ucsd.edu).
	       This material is based upon work supported by the National Science Foundation under grant nos. NSF-ECCS-2153698, NSF-CCF-2225555, NSF-CNS-2147955  and is supported in part by funds from federal agency and industry partners as specified in the Resilient \& Intelligent NextG Systems (RINGS) program.}
}

\maketitle
\bstctlcite{IEEEexample:BSTcontrol}

\begin{abstract}
    Obtaining accurate and timely channel state information (CSI) is a fundamental challenge for large antenna systems. Mobile systems like 5G use a beam management framework that joins the initial access, beamforming, CSI acquisition, and data transmission. The design of codebooks for these stages, however, is challenging due to their interrelationships, varying array sizes, and site-specific channel and user distributions. Furthermore, beam management is often focused on single-sector operations while ignoring the overarching network- and system-level optimization. In this paper, we proposed an end-to-end learned codebook design algorithm, network beamspace learning (NBL), that captures and optimizes codebooks to mitigate interference while maximizing the achievable performance with extremely large hybrid arrays. The proposed algorithm requires limited shared information yet designs codebooks that outperform traditional codebooks by over $10$dB in beam alignment and achieve more than $25\%$ improvements in network spectral efficiency.
\end{abstract}


\section{Introduction}
    Multiple-input multiple-output (MIMO) systems have played a critical role in supporting high-speed wireless communications \cite{HeathLozano2018}. Between spatial multiplexing and beamforming, array-based systems have supported new verticals and applications since 3GPP Release $7$ \cite{Dreifuerst2023magazine}. 
    Beam management is an important component of expanding MIMO to massive MIMO and extreme MIMO (X-MIMO) \cite{Holma2021XMIMO}.
    Beam management employs codebook-based strategies for beam training with hybrid arrays \cite{HeathOverviewSP4mmWave2016, Hamid2023Hybridsurvey} and configurable feedback \cite{Dreifuerst2023magazine, Giordani3GPPBeamManagement2019} to support large-scale arrays. While beam management enables improved flexibility and higher dimensional MIMO arrays, the performance is dependent on the beam codebooks employed throughout the process. Yet optimizing codebooks for beam management is challenging because each codebook is dependent on the other system processes. As a result of this lack of optimization, massive MIMO, and even multi-user MIMO deployments are still limited in 5G \cite{Dreifuerst2023hierarchical}.

    Beam management, from a system level, is a series of operations enabling initial access, beam refinement, and feedback that ultimately lead to data transmission \cite{Dreifuerst2023hierarchical, Giordani3GPPBeamManagement2019}. Beam tracking and other beam search reduction techniques can also be understood within the overarching beam management framework. The primary operations can be referenced by the codebooks and 3GPP signals transmitted during the respective stages. First, user equipment (UE) joins and synchronizes with the network using a pre-defined synchronization signal block (SSB) sequence that is beamformed using a beamforming codeword from a codebook stylized as the \textit{SSB codebook} \cite{Dreifuerst2022VTCBM}. Broadly speaking, this beamforming is intended to ensure all users can accurately synchronize, requires a small amount of time-frequency resources, and trades off beamforming gain for broad coverage. In contrast, beam refinement requires a large amount of resources to enable channel estimation on top of a fine-grained beam search using another beamforming codebook. Channel state information reference signals (CSI-RS) are transmitted using a subset of the \textit{CSI-RS codebook} based on the user-reported SSB feedback. The beamformed channel represents a compressed channel and allows for reduced-rank feedback and increased coherence time \cite{V.VaEtAlImpactBeamwidthTemporalChannel2017}. Finally, a third codebook (referred to as the \textit{feedback codebook}) is used to quantize the beamformed channel estimates to reduce the overhead of feedback signaling with large arrays. The feedback codebook is configurable and standardized in 5G, with the most relevant format for MU-MIMO (type-II) defined as an oversampled DFT codebook \cite{Qin2023CSIFeedback}. Using the combination of coarse and refined beam training as well as quantized feedback, a serving cell can design a set of precoders to serve the UEs for data transmission. 

    Although beam management has been standardized since 3GPP Release $15$ \cite{Dreifuerst2023magazine}, new research for optimizing beam management to improve the performance in 6G is underway \cite{Lin2023OverviewAI5G}.
    At the forefront of this innovation are extensions and improvements for beam management including support for 128 ports \cite{3gpp128pMIMO}. Yet, configuring, optimizing, and updating codebooks is challenging because mathematically capturing the codebook interrelationships in addition to user distributions and realistic channel models is often intractable or imprecise. As a result, 3GPP has identified beam management and CSI as potential areas to integrate artificial intelligence/machine learning (AI/ML) \cite{Lin2023OverviewAI5G}. These data-driven techniques present new possibilities for designing site-specific algorithms without explicitly modeling distributions within the data \cite{JiangML46GWirelessSurvey2017, ZhangDLinWirelessSurvey2019, HengSixChallengesBM6G2021}. Within the scope of AI/ML for beam management, we identify three directions of work relevant to our investigation--codebook design, beam training, and interference mitigation. 

    \textbf{Prior work on codebook design:}
    AI/ML is used in the design or implementation of beamforming codebooks \cite{ShafinEtAlDRL4BeamOptim2020, HengProbingBeams2021, Dreifuerst2022VTCBM} or CSI feedback codebooks \cite{Turan2024LimitedGMM, Chen2023CSInet, Lee2023DenoisingCSI, Ma2023DLCSI}. Although beamforming and feedback codebooks are used for different roles, the systems and techniques for applying AI/ML to these problems are very similar. For beamforming codebooks, hierarchical codebooks have been a primary focus and adopted for 5G since early proposals \cite{Wangetal2009CodebookTraining, Alkhateeb2014, Morozov2016}. Prior work has considered supervised learning \cite{HengProbingBeams2021, Dreifuerst2022VTCBM, Yang2023hierarchicalBeamAlignment} and reinforcement learning \cite{ShafinEtAlDRL4BeamOptim2020} styles of learning initial access codebooks. That work (\cite{HengProbingBeams2021, Dreifuerst2022VTCBM, Yang2023hierarchicalBeamAlignment, ShafinEtAlDRL4BeamOptim2020}) does not consider how the new beamforming codebooks impact beamformed channel feedback nor the achievable rate in MU-MIMO data transmission. Ignoring the system-level impact of codebook design is a reoccurring result in prior work on beam management where a subset of the process--either beam training or codebook design--is modified but the overarching metrics are not considered. 
    
    Codebook design using AI/ML has also been explored for feedback quantization and reconstruction \cite{WenDL4mMIMOCSIFeedback2018, Chen2021, Lee2023DenoisingCSI, Turan2024LimitedGMM, Chen2023CSInet, Xiao2023MetaCSI}. Deep learning autoencoder models have been used to replace feedback codebooks \cite{WenDL4mMIMOCSIFeedback2018, Chen2021} or as a denoising process to improve the received feedback \cite{Lee2023DenoisingCSI, Turan2024LimitedGMM, Chen2023CSInet, Xiao2023MetaCSI}. Feedback autoencoders require the encoder and decoder of the model to be trained together and then shared throughout the network. While sharing a neural network may be reasonable for higher-level tasks, the size of the models and difficulty with standardizing these models make them ill-suited for practical deployments. Denoising algorithms that operate in a one-sided fashion \cite{Lee2023DenoisingCSI, Turan2024LimitedGMM, Chen2023CSInet} are advantageous as they can be employed at the BS and improve the uplink feedback, which typically operates at a low SNR due to uplink power constraints \cite{Bjornsson2022CombiningCSIFeedback}. The challenge with denoising algorithms is they tend to be site-specific, not generalize well, and are not applicable for dynamic beamformed channels as seen by hybrid arrays \cite{Dreifuerst2023hierarchical}. Some work on generalization or meta-learning has been proposed \cite{Xiao2023MetaCSI} though it still ignores the prior beam management steps that impact the effective channel and CSI distribution.

    \textbf{Prior work on beam training:}
    The second direction of focus on AI/ML in beam management is beam training, which is the process of determining the beam-codewords to use from the standard beamforming codebooks. Beam training has received significant attention toward the goal of maximizing the wireless link power with limited overhead \cite{Yang2022DLBeamAlignment, Sim2018MABBMV2X, Wangetal2009CodebookTraining, HeathetAlVehicularSensing2016, V.VaEtAlInverseMultipathFingerprintingMillimeter2018, Hussain2022DualtimeBeamTracking}. Data-driven techniques have been used to integrate localization and sensing information \cite{HeathetAlVehicularSensing2016}, capture site-specific characteristics \cite{Yang2023hierarchicalBeamAlignment}, or improve mobility robustness \cite{V.VaEtAlInverseMultipathFingerprintingMillimeter2018, rebato2018multiTRP, Hussain2022DualtimeBeamTracking}. Additional work on a generalized beam selection algorithm was proposed in \cite{Vuckovic2023GeneralBeamSel}. These investigations are focused on mmWave bands where the environment is sparse. While sparse environments can challenge beam training, the sparsity also results in the beam search primarily focused on finding a rank-$1$ beamformer aligned with a single path. Lower bands such as sub-6GHz or upper-mid bands \cite{Holma2021XMIMO} exhibit richer scattering that makes the beam training process depend on coherent combining over multiple paths and limiting interference instead of simply finding a main beamforming direction. Instead, these works \cite{Yang2022DLBeamAlignment, Sim2018MABBMV2X, Wangetal2009CodebookTraining, HeathetAlVehicularSensing2016, V.VaEtAlInverseMultipathFingerprintingMillimeter2018, Hussain2022DualtimeBeamTracking} are primarily targeting analog beamforming and do not consider MIMO formats. This is a significant limitation in mobile broadband, which has supported SU-MIMO and digital systems since 3GPP release $7$ \cite{Dreifuerst2023magazine}. 

    \textbf{Prior work on codebook-based interference mitigation:}
    Interference-aware beam management is a combination of codebook design and beam training with additional consideration for how overlapping beams cause interference in adjacent sectors and cells. In one line of work, a conventional beam management interference cancellation algorithm was converted to a deep learning approximate to reduce complexity \cite{Zhou2019Interference}. A direct AI/ML algorithm for interference-aware codebooks was also recently proposed \cite{Zhang2024interferenceBM}. These algorithms (\cite{Zhou2019Interference, Zhang2024interferenceBM}), though, are trained on data from mmWave bands and therefore benefit from the sparse channel nature while also only considering simplified SU-MISO systems. There is a lack of research investigating data-driven codebook algorithms that operate in rich scattering environments and in MU-MIMO settings.

    \textbf{Prior work on beamspace codebook design:} Our prior work introduced the concepts of beamspace codebook learning and hierarchical codebook learning for X-MIMO arrays \cite{Dreifuerst2023hierarchical, Dreifuerst2023CodebookFeedback}. We proposed a learning algorithm that produced highly effective beam training codebooks but only provided a modest improvement in the overall spectral efficiency. In our present work, we extend the end-to-end training to encapsulate the achievable spectral efficiency, thereby producing codebooks that support full rank MIMO connections, while also mitigating inter-cell interference. Contributions of this paper:
   \begin{itemize}
        \item We describe a complete beam management process for a multi-cell, hybrid, X-MIMO system. The beam process includes inter- and intra-cell interference during beam training and data transmission to characterize the impact of interference with dynamic codebooks. By capturing and modeling each stage from initial access through network spectral efficiency, we can obtain insights into the net effects of advanced codebook algorithms in realistic settings.
        
        \item We propose an end-to-end learning strategy and neural processing architecture ``Network Beamspace Learning'' ($\algo$) to design SSB and CSI-RS codebooks for multiple base stations serving overlapping regions. The strategy is based on representing beamformers in the beamspace and backpropagating gradients through the entire beam management model from initial access through achievable rate calculation. The end-to-end training strategy ensures that codebooks are learned to maximize the beamformed channel and mitigate interference that would reduce spectral efficiency. The $\algo$ formulation only requires sharing the SSB feedback, yet significantly improves the overall system performance.
        
        \item We evaluate the capabilities of the beam management framework with X-MIMO using traditional and $\algo$ codebooks across a range of possible deployments. In particular, we evaluate using different array geometries and codebook sizes in raytraced environments. Ultimately, we find that MU-MIMO systems are susceptible to interference with extremely large arrays, but learned codebooks can mitigate the interference and achieve significantly higher cell spectral efficiency.
    \end{itemize}   

    \begin{figure*}[!t]
        \centering
        \includegraphics[width=6.25in]{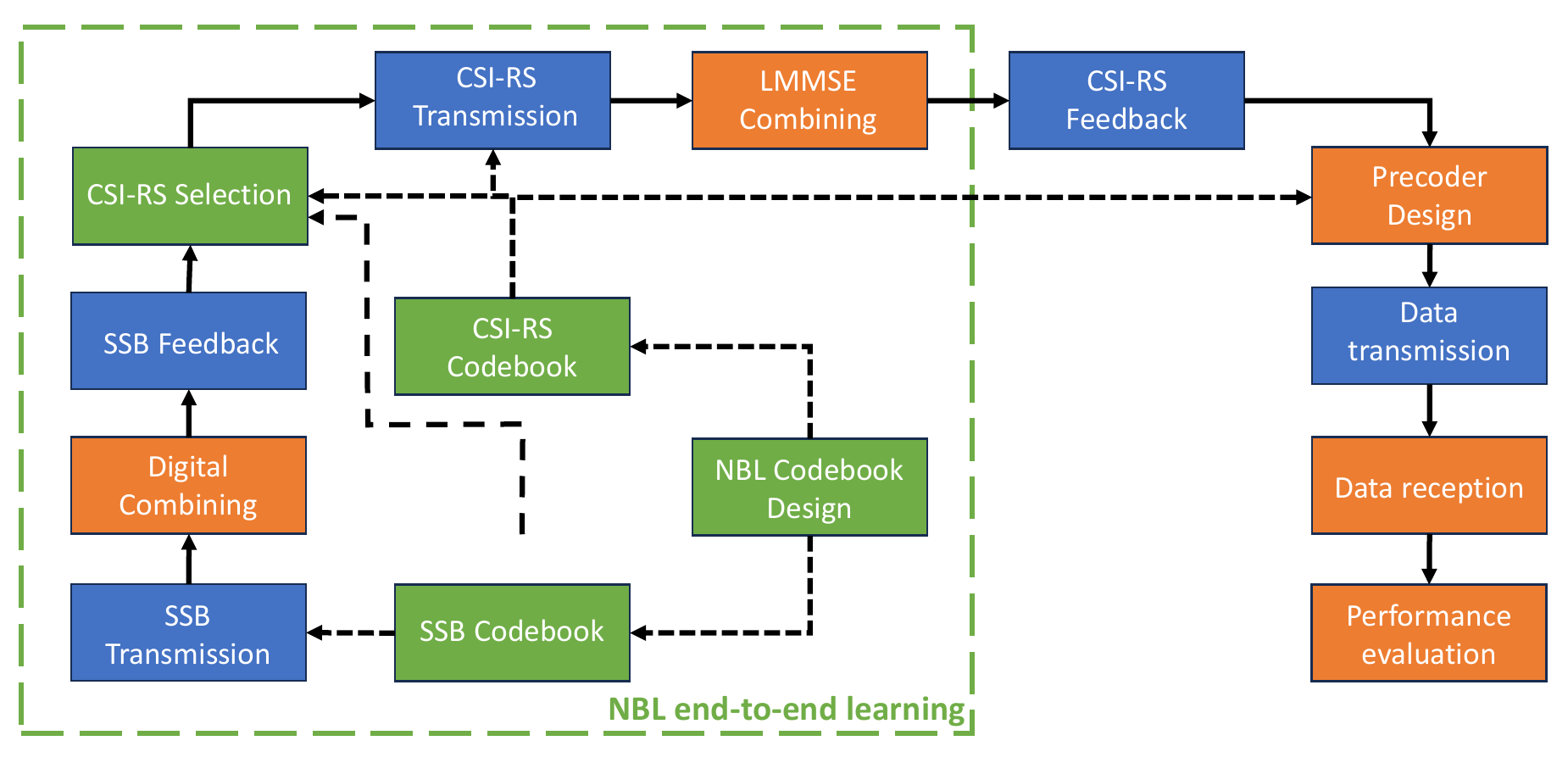}
        \caption{A block diagram of the system-level model and $\algo$ codebook roles. The proposed solution modifies the green blocks, while the orange and blue blocks refer to assumed and standard operations respectively. The end-to-end learning framework is shown by the green dashed box that encapsulates the codebook operations. The assumed operations outside of the end-to-end learning are not critical but allow for system-level evaluation.}
        \label{fig: block_diagram}
    \end{figure*}

    $\mathbf{Notation}$: $\vect{A}$ is a matrix, $\vect{a}$ and $\{a[i]\}_{i=1}^{N}$ are column vectors and $a,A$ denote scalars. $\vect{A}^T$, $\overline{\vect{A}}$, $\vect{A}^*$, and $\vect{A}^{\dagger}$ represent the transpose, conjugate, conjugate transpose, and psuedo-inverse of $\vect{A}$. 
    $\vect{A}[k, \ell]$ denotes the entry of $\vect{A}$ in the $k^{\text{th}}$ row and the $\ell^{\text{th}}$ column. The same meaning is also associated with $\vect{A}_{k, \ell}$. Similarly, $\vect{A}[:, k]$ refers to the $k^\text{th}$ column of $\vect{A}$. Unspecified norm equations are $\norm{\mathbf{a}}_{2} = \mathbf{a}^* \mathbf{a}$ for vectors and the Frobenius norm $\norm{\mathbf{A}}_F = \sqrt{\text{Tr}(\mathbf{A} \vect{A}^*)}$ for matrices. The operator $\mathbb{E}[\cdot]$ is used for the expectation of a random variable. Due to the notational complexity of MU-MIMO with OFDM, we will always use $c$ and $u$ to refer to a specific base station cell and UE, $t$ as a specific time, $k$ as a specific frequency resource, and $\nt / \nr$ to refer to a specific transmit or receive antenna.
    We use the superscript $\analog{(\cdot)}$ to refer to the analog component of hybrid variables within the MIMO system.

    The remainder of our paper is organized as follows. 
    First, we introduce the multi-cell beam management model that includes the SSB and CSI-RS processes as depicted by the block diagram in Figure \ref{fig: block_diagram}. We carefully define the transmission and reception of these signals as these expressions are directly integrated into the $\algo$ backpropagation. Then, we describe a typical feedback and data transmission strategy that we use for evaluating the system-level performance of the proposed codebook algorithm in Sectoin \ref{sec: feedback}. In Section \ref{sec: algo} we describe $\algo$ which includes beamspace preprocessing, a multi-stage neural network, and achievable-rate maximizing end-to-end learning. In the final sections, we present the simulation setup and a rigorous evaluation of the various codebooks, overall performance, and out-of-distribution (OOD) performance achieved by our proposed method. 


\section{Multi-cell beam management} \label{sec: system}
    In this section, we define the hierarchical beam training operation for a multi-cell scenario. Although beam training is standardized in 5G, we explicitly define each step of the beam-based operations because the multi-cell beam management process is integrated into the gradient updates of the learning algorithm in Section \ref{sec: algo}. 

    We consider a region that is served by $C=3$ base stations (BS) or cells with overlapping coverage regions. Exterior cells are not simulated in this investigation, although extending the model and algorithm to include more BSs is straightforward. Each BS is equipped with a planar array of size $\Nta = 2\Nxa\times \Nya$ with 2 horizontal elements of perpendicular polarization being colocated in the array. The antenna elements are connected in a fully connected hybrid format with $\phaseres$ bit phase shifters connecting to the $2\Nx\times\Ny$ digital ports \cite{Alouzi2023FullHybridArray}. We assume there are $U$ UEs in the scene where $U$ is a random variable that is not known to the network. Each UE is equipped with an $\Nr$ element fully digital uniform linear array enabling elevation beamforming. We assume a multi-cell MU-MIMO OFDM channels between each antenna pair as
    \begin{equation}
        \Achan \in \complex^{C \times U \times T \times K \times \Nr \times \Nta} \label{eqn: Achan}.
    \end{equation}
    The specific channel model is not critical as underlying distributions of the channel are learned from channel samples.

    For a realized channel, we can define a received signal model for a user $u$ at time-frequency resource $t, k$ with receiver combiner $\tku{\vect{W}^*}$ and hybrid precoders $\vect{F}_{c, t, k}$, $\Fa$. We will assume timing and frequency synchronization can be performed using standardized reference signals. We also assume all users can be associated with a BS so that the U users can be partitioned $U = \sum_c U_c$ for ease of notation, although the user assignment may not be known prior to initial access or shared between cells. We define the received signals as
    \begin{align}
        \tku{\vect{D}} &= \vect{H}_{c, u, t, k} \Fa \F_{c, t, k} \vect{s}_{c, t, k} \\
        \vect{I}^{\text{intra}}_{c, u, t, k} &= \sum_{c'\ne c}^C\vect{H}_{c', u, t, k} \Fap \vect{F}_{c', t, k} \vect{s}_{c', t, k} \label{eqn: intra}\\
        \tku{\vect{y}} =& \tku{\vect{W}^*}  \left(\vect{D}_{u, t, k} + \vect{I}^{\text{intra}}_{c, u, t, k} + \vect{N}_{u, t, k}\right). \label{eqn: rec_signal}
    \end{align}
    The noise $\tku{\vect{N}}$ is modeled as independent Gaussian random values and accounts for thermal noise and hardware noise figures. Note that $\tku{\vect{D}}$ contains the desired signal for a given user and the inter-cell interference from streams of data for other users, while the intra-cell interference is given by \eqref{eqn: intra}. The digital precoders $\F_{c, t, k} \in \complex^{\Nt\times \Nr}$ are block diagonal where a subset of the digital array is used to serve and mitigate interference for each user $u\in U_c$ \cite{Dreifuerst2023hierarchical}. The analog precoders $\Fa \in \complex^{\Nta \times \Nt}$ are a matrix of frequency-flat beamformers used to direct the data streams. 

    The task of determining analog and digital precorders that maximize the network-wide spectral efficiency (SE) is challenging in the face of limited CSI and coordination. Beam management is one strategy that involves multiple stages of beam training and feedback to obtain CSI and configure the precoders. In particular, the beam training stages are used to obtain coarse and refined beamformers that make up the analog beamforming, while user feedback is used to configure the digital precoding \cite{Giordani3GPPBeamManagement2019, Dreifuerst2023hierarchical}. The focus of our work here is on optimizing network performance by dynamically generating new beamforming codebooks. 

    For the duration of our paper, we assume a working knowledge of the beam management stages and will only briefly highlight the key equations. Readers are encouraged to review \cite{Dreifuerst2023magazine, Dreifuerst2023CodebookFeedback, Giordani3GPPBeamManagement2019, Dreifuerst2023hierarchical} for an in-depth review of these processes in 5G. Given the focus on beam management for 6G \cite{Lin2023OverviewAI5G}, we will assume the same three phases (SSB, CSI-RS, and feedback) as in 5G. We also assume UEs operate similarly to sub-6GHz 5G implementations, although such assumptions are just a simplification to focus on the BS operation. Much of the beam training and hybrid operations are not specified in the standards, so we will assume a formulation similar to classical hybrid implementations found in literature i.e. \cite{HeathOverviewSP4mmWave2016, Singh2021SurveyHybridBF, Wu2018HybridMUMIMO} and following the unified hybrid beam management system model from our prior work \cite{Dreifuerst2023hierarchical}. One important difference is the multi-cell formulation, where we will assume the \emph{nearby cells operate over the same bands and perform beam training on the same time-frequency resources}. This assumption is a worst-case scenario as nearby base stations could employ different bands for SSB and CSI-RS to mitigate multi-cell interference.

    \subsection{Synchronization signal block transmission}
        Initial access begins by transmitting a small number of broadcast beams with synchronization and basic network configuration information \cite{Giordani3GPPBeamManagement2019, Dreifuerst2023magazine}. The process of transmitting a burst of SSBs is repeated periodically and allows new users to join the network. Historically, SSB codebooks were simple codebooks designed to form wide beams to cover large areas with limited beamforming gain \cite{Raghavan2016BFIA}. Directional beamforming, however, is not well-suited for multipath environments as experienced in sub-6GHz and frequency range (FR) $1$ bands \cite{Dreifuerst2023CodebookFeedback}. With the upcoming increased support for X-MIMO arrays expected in 6G, new codebooks that operate in rich scattering environments and across array sizes are critical.

        The SSB signals are transmitted on frequency resources $\Kssb$ and specified timeslots $\Tssb$ for each SSB $i$ up to the total SSB size $\Lmax$. The received SSB signal before combining can be simplified from \eqref{eqn: rec_signal} to a single-stream representation of the demodulation reference signals $s^{\text{DMRS}}_{c, t, k} \in \complex$ as
        \begin{align}
            \vect{y}^{\text{SSB}_i}_{c, \ue, t, k} =\ &\frac{1}{\sqrt{K \Nt}} \Achan_{c, u, t, k} \f_{c, i} s^{\text{DMRS}}_{c, t, k} \\
            & + \sum_{c'\ne c}^C\vect{H}_{c', u, t, k} \f_{c', i} s^{\text{DMRS}}_{c', t, k} + \vect{n}_{u, t, k} \label{eqn: ssb}
        \end{align}
        where $\f_{c, i} \in \Bssb_{c}$ is the analog beamformer controlled from a single digital port. Effectively, a digital beamformer could also be used to compensate for low-resolution phase shifters in the hybrid array, although the effects are almost imperceptible during rank-$1$ beamforming \cite{Alouzi2023FullHybridArray, Dreifuerst2023hierarchical}. We will assume the UE performs digital combining for each cell-specific reference signal to maximize the SSB RSRP as
        \begin{align}
            \RSRPciu &=\ \sum_{k\in \Kssb} \sum_{t \in \Tssb} \norm{\vect{y}^{\text{SSB}_i}_{c, \ue, t, k}}^2. \label{eqn: RSRP}
        \end{align}
        In general, UEs could be configured for beam training or other combining strategies, but our focus is on the base station operations so we assume digital combining for simplicity. 

        After SSB reception, the UEs provide a short feedback packet that includes a selected base station, SSB index, and corresponding RSRP obtained during the selected SSB. In total, the aggregate feedback in the network is represented as
        \begin{align}
                \vect{p} &=\ \left\{\max_i{\RSRPciu} \right\}^{U}_{\ue=1} \label{eqn: Prsrp} \\
                \vect{b}, \vect{m} &=\ \left\{\argmax_{c, i}{\RSRPciu} \right\}^{U}_{\ue=1}. \label{eqn: Irsrp}
        \end{align}
        The vectors $\vect{m}$ and $\vect{b}$ correspond to the best cell-beam index \eqref{eqn: Irsrp} and $\vect{p}$ is the corresponding RSRP \eqref{eqn: Prsrp} for all $U$ users. Note that $\vect{p}$ would be quantized in realistic systems, although the quantization resolution is $\le1$dB so we neglect the RSRP quantization for this investigation. We will formulate the problem around a weakly centralized controller that obtains the slowly-updated feedback from \eqref{eqn: Prsrp}-\eqref{eqn: Irsrp}. In general, however, each cell may only have access to the information of users for its cell, especially for the CSI-RS step following SSB feedback. The cell-specific feedback is denoted
        \begin{align}
                \vect{p}^{(c)} &=\ \left\{\max_i{\RSRPciu} \right\}_{\ue\in U_c} \\
                \vect{m}^{(c)}, \vect{b}^{(c)} &=\ \left\{\argmax_{c, i}{\RSRPciu} \right\}_{\ue\in U_c}.
        \end{align}
        After the SSB feedback, each BS can transmit CSI-RS to obtain more accurate beam alignment and channel estimates in the form of PMI.

    \subsection{CSI-RS transmission}
        Refined beam training is performed by transmitting CSI-RS using a hierarchical codebook strategy following SSB feedback. CSI-RS occupy significantly larger resources and enable pilot-based channel estimation. Typical CSI-RS deployments begin with a large CSI-RS codebook ($\gg\Lmax$) and then perform a subset selection based on the SSB feedback \cite{Dreifuerst2023CodebookFeedback}. We define $\Bactive$ as the active CSI-RS codebook subset and employ a selection algorithm similar to \cite{Dreifuerst2023hierarchical}. The important difference is instead of searching over each vector, the search is performed over precoder matrices, and the most correlated vectors from each precoder are compared for subset selection. The benefit to precoder-based selection is that the goal is no longer to achieve a set of rank-$1$ beamformers and then spatially multiplex with them independently; instead, the codebook contains sets of precoders that are likely to support multi-stream communication. In particular, we modify \cite[equation ($8$)]{Dreifuerst2023hierarchical} to instead be
        \begin{align}
            \vect{C} = \max_s\left<\Bssb_{c, i}, \Bcsi_{c, j, s}\right> \quad \forall i,j. \label{alg: prop_sel1}
        \end{align}
        From there, each step is essentially identical with the selected codewords now corresponding to matrices instead of grouping the beamforming vectors in the following step. The CSI-RS subset selection process is repeated for each cell independently using the cell-specific feedback $\vect{p}^{(c)}, \vect{m}^{(c)}, \vect{b}^{(c)}$.
        The result is a subset of $\Ncsi$ CSI-RS beams for each base station that is proportionally selected to have at least one beamformer that is highly correlated with the active SSB beams reported.

        \begin{figure}[!t]
            \centering
            \includegraphics[width=3.25in]{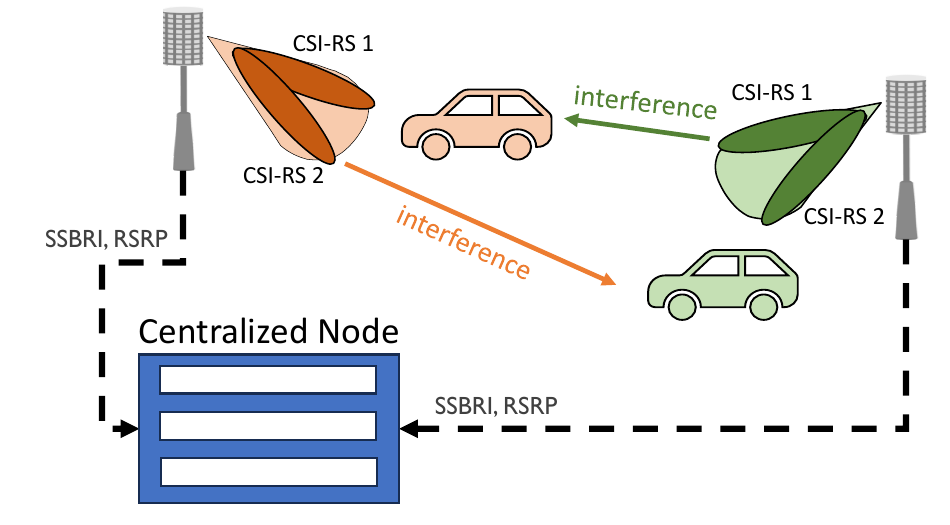}
            \caption{Users are allocated to the cells during SSB transmission. During CSI-RS transmission, only interference during a UE's selected CSI-RS causes interference. Therefore, codebooks should be designed so that the first green CSI-RS does not interfere strongly with the orange UE and vice versa.}
            \label{fig: system_beams}
        \end{figure}
            
        The cells begin the CSI-RS transmission process by transmitting training pilot symbols $\Str\in \complex^{\Bg \times 1}$ with the active subset of beamformers $\Bactive_{c}$. The received CSI-RS signal for each UE before combining is then
        \begin{align}
            \ycsirs_{\ue, t, k} =\ &\Achan_{\vect{b}_u, u, t, k} \Bactive_{\vect{b}_u, i} \vect{s}^{\text{tr}}_{\vect{b}_u, t, k}\\ &+ \sum_{b'\ne\vect{b}_u} \Achan_{b', u, t, k} \Bactive_{b', i} \vect{s}^{\text{tr}}_{b', t, k} + \vect{n}_{u, t, k}.
        \end{align}
        Importantly, the users are now associated with a given base station $\vect{b}_u$ determined during the SSB state. Therefore, all other cells create interference as a result of their CSI-RS transmissions in the same time-frequency blocks as shown in Figure \ref{fig: system_beams}. The UE, however, can utilize its antenna array to combine the intended signal and mitigate interference, for example, with a linear minimum mean squared error (LMMSE) receive combiner \cite[Section 9.4]{HeathLozano2018} resulting in
        \begin{align}
            \vect{R}_{u, t, k} &= \left(\sum_{c}^C \Achan_{c, u, t, k} \Bactive_{c, i} ( \Achan_{c, u, t, k} \Bactive_{c, i})^* \right) \\
            \vect{V}_{i, u, t, k} &\triangleq \Achan_{b_u, u, t, k} \Bactive_{b_u, i} \\
            \vect{W}^{\text{CSI-RS}_i}_{u, t, k} &= \vect{R}_{u, t, k}^{-1} \vect{V}_{i, u, t, k} \label{eqn: LMMSE}\\
            s^{\text{CSI-RS}_i}_{u, t, k, \nr} &= \frac{ (\vect{V}_{i, u, t, k})^*_{\nr} \vect{R}_{u, t, k}^{-1} (\vect{V}_{i, u, t, k})_{\nr}  }{1-    (\vect{V}_{i, u, t, k})^*_{\nr} \vect{R}_{u, t, k}^{-1} (\vect{V}_{i, u, t, k})_{\nr}}. \label{eqn: SINR}
        \end{align}
        Equation \eqref{eqn: SINR} calculates the signal-to-interference-noise ratio (SINR) based on receiver combiner $\vect{W}^{\text{CSI-RS}_i}_{u, t, k}$ in \eqref{eqn: LMMSE}. The SINR is a configurable reporting metric in 5G  and is measured using zero-power and non-zero-power CSI-RS \cite{Dreifuerst2023magazine}. 
        
        Finally, the UE selects the CSI-RS resource that would maximize the rate based on a configurable metric estimated from the pilot symbols. The two standardized metrics are SINR or RSRP, although SINR is preferred for interference management. Mathematically the selection process is based on the average over the time-frequency resource ($\Tcsi, \Kcsi$)
        \begin{align}
            \text{SE}^{\text{CSI-RS}_i}_{u} &= \sum_{\nr=1}^{\Nr}\log_2 \left(1 + \frac{1}{\Tcsi \Kcsi} \sum_{t, k} s^{\text{CSI-RS}_i}_{u, t, k, \nr}  \right) \label{eqn: ach_rate}\\
            \ihat &= \left\{\argmax_{i} \text{SE}^{\text{CSI-RS}_i}_{u}\right\}_{u\in U^{c}}.
        \end{align}
        The analog beam selections, $\ihat$ correspond to the chosen analog precoders that will be used by the base station during hybrid data transmission because all of the UE feedback is based on the beamformed channels, $\Achan_{b_u, u, t, k} \Bactive_{b_u, \widehat{i}_{b_u, u}}$.
                    
        Compared to the SSB model \eqref{eqn: ssb}, the CSI-RS received model is multi-layer due to the need for sounding each of the $\Bg$ ports for each user. 
        

    \section{System model} \label{sec: feedback}
        In this section, we define a typical implementation of feedback and data transmission using the CSI available through beam management. While the previous section is integrated within the $\algo$ learning process, this section can be understood as the ``system model'' surrounding our algorithm that allows for evaluating the proposed algorithm within a realistic system. The system model can be thought of as modular within wireless systems, where the algorithms for feedback or data transmission can be replaced without impacting the proposed solution.

    \subsection{Feedback}
        The SINR \eqref{eqn: SINR} impacts the channel estimation efficacy and is useful for determining the modulation and coding scheme (MCS) for data transmission. The beamformed channel estimate is defined for a channel estimation algorithm $f(\cdot)$ as
        \begin{align}
            \widehat{\vect{H}}_{u, t, k} &\triangleq f(\ycsirs_{u}, \Str)
        \end{align}
        For ease of notation, we define $\widehat{\vect{H}}_{u, t, k}$ as the beamformed channel estimate--rather than the physical channel estimate--with the assigned base station. To be precise, $\widehat{\vect{H}}_{u, t, k}$ represents an estimate of the beamformed channel $\vect{H}_{b_u, u, t, k}\vect{F}_{b_u, i_{b_u, u}}$. 
        There is no standardized channel estimation algorithm, but, given the limited CSI-RS response timing and power budget of mobile UEs, an LMMSE method is assumed herein. The resulting channel estimate at the UE is frequency selective over a set of frequency subbands $b \in \{1, ... \bwp\}$ and carried out in a similar manner to \cite{Dreifuerst2023CodebookFeedback}. For ease of notation, we focus on a single subband although employ multiple subbands in the simulation results in Section \ref{sec: Sim_results}. In the final step, the channel estimates are quantized according to a feedback codebook and transmitted to the BS \cite{Dreifuerst2023magazine, Qin2023CSIFeedback}.

       The signal-to-interference-noise ratio (SINR) presented in \eqref{eqn: SINR} provides a valuable secondary metric for the beam management efficacy as it ultimately determines the channel estimation performance as a result of beam alignment with the desired serving cell and interference from nearby cells. In Section \ref{sec: algo} we describe how we use the SINR and achievable spectral efficiency \eqref{eqn: ach_rate} resulting from the entire beam management process in the end-to-end learning framework.

    \subsection{Data transmission}
        Following the first three stages of beam management (SSB, CSI-RS, and feedback), the network can configure and transmit data using the combined analog precoders $\Fistar$ and corresponding feedback $\widehat{\vect{H}}_{u}$. In particular, we formulate the precoder design algorithm in the same way as traditional hybrid formulations to make use of well-known techniques \cite{HeathLozano2018, HeathOverviewSP4mmWave2016, Singh2021SurveyHybridBF}. Conceptually, the analog beams chosen from the CSI-RS codebook provide beamforming gain, while a regularized zero-forcing precoder (RZF) is used to configure the digital precoding to mitigate intracell interference. Furthermore, each cell does not have feedback for users that are associated with other cells, so each cell operates independently of the other cells at this stage. Therefore, the following steps directly follow the single-cell RZF formulation from \cite{Dreifuerst2023hierarchical} with the key equations included here for reference.

        RZF precoding is based on maximizing the signal-to-leakage noise ratio, thereby allowing for some inter-stream interference to improve low-SNR performance \cite[section 9.9]{HeathLozano2018}. Without loss of generality, we assume the users are ordered so that the first $U^{\text{a}}_c$ users are actively selected for joint transmission according to some scheduling algorithm. Then the analog precoding is defined as
        \begin{align}
            \analog{\F}_{c} &= [\F_{c, \hat{i}_0}, \F_{c, \hat{i}_1}, ..., \F_{c, \hat{i}_{U^{\text{a}}_c}} ]. \label{eqn: F_RF}
        \end{align}
        The digital precoders are determined for each user and associated in a block diagonal form based on the rank of the CSI-RS $\Bg$
        \begin{align}
              \vect{F}_{c, u, t, k} &= \frac{(\sum_{i=0}^{U^{\text{a}}_c-1} \widehat{\vect{H}}_{i, t, k}^*\widehat{\vect{H}}_{i, t, k} + U \Nt\mathbb{E}[\vect{N}_{u, t, k}^2])^{-1} \widehat{\vect{H}}_{u, t, k}^*}{\norm{(\sum_{i=0}^{U^{\text{a}}_c-1} \widehat{\vect{H}}_{i, t, k}^*\widehat{\vect{H}}_{i, t, k} + U \Nt\mathbb{E}[\vect{N}_{u, t, k}^2])^{-1} \widehat{\vect{H}}_{u, t, k}^*}}\\
             \F_{c, t, k} &= 
             \begin{bmatrix}
                \vect{F}_{c, 0, t, k} & 0 & ... & 0  \\
                0 & \vect{F}_{c, 1, t, k} & 0 & ... \\
                \vdots & \vdots & \ddots & \vdots\\
                0 & 0 & ... & \vect{F}_{c, U^{\text{a}}_c, t, k} 
             \end{bmatrix}.  \label{eqn: digital_precoding}
        \end{align}
        With the precoding defined, we can finalize the steps necessary to reach the network metric: effective sum spectral efficiency (ESSE). We assume the same LMMSE receive combining strategy by the UEs as defined in \eqref{eqn: LMMSE}, with the slight modification for the data precoders instead of the CSI-RS precoders. Then the per-user SINR is direct copy of \eqref{eqn: SINR} with the replaced variables. Finally, the network-wide sum spectral efficiency after removing the time-frequency resources used for beam management ($T_{\text{BM}}, K_{\text{BM}}$) is
        \begin{align}
            \text{ESSE} &= \sum^{U}_{u=1} \sum_{\substack{t \notin T_{\text{BM}} \\ k\notin K_{\text{BM}}}} \sum_{r=1}^{R} \log_2 (1 + s_{u, t, k, r}). \label{eqn: ESSE}
        \end{align}
        Note that the beam management process up through data transmission is repeated regularly throughout the wireless network. During each period, the codebooks can be retained and new codebooks can be generated to best serve the specific site characteristics including user dynamics and environment characteristics. The periodic nature of beam management is the basis of our proposed algorithm, $\algo$, which employs deep learning to generate codebooks between beam management periods based on the limited feedback and prior codebook information. By employing deep learning to generate codebooks, as opposed to channel estimation or precoder design, the system can circumvent many of the timing and compute constraints of deep learning when it comes to wireless communications. In addition, by explicitly modeling the beam management steps, a neural network can be trained with gradients calculated through the entire process in an end-to-end style \cite{Hoydis2018E2ELearning}.

\section{Network beamspace learning} \label{sec: algo}
\begin{figure*}[!t]
        \centering
        \includegraphics[width=\textwidth]{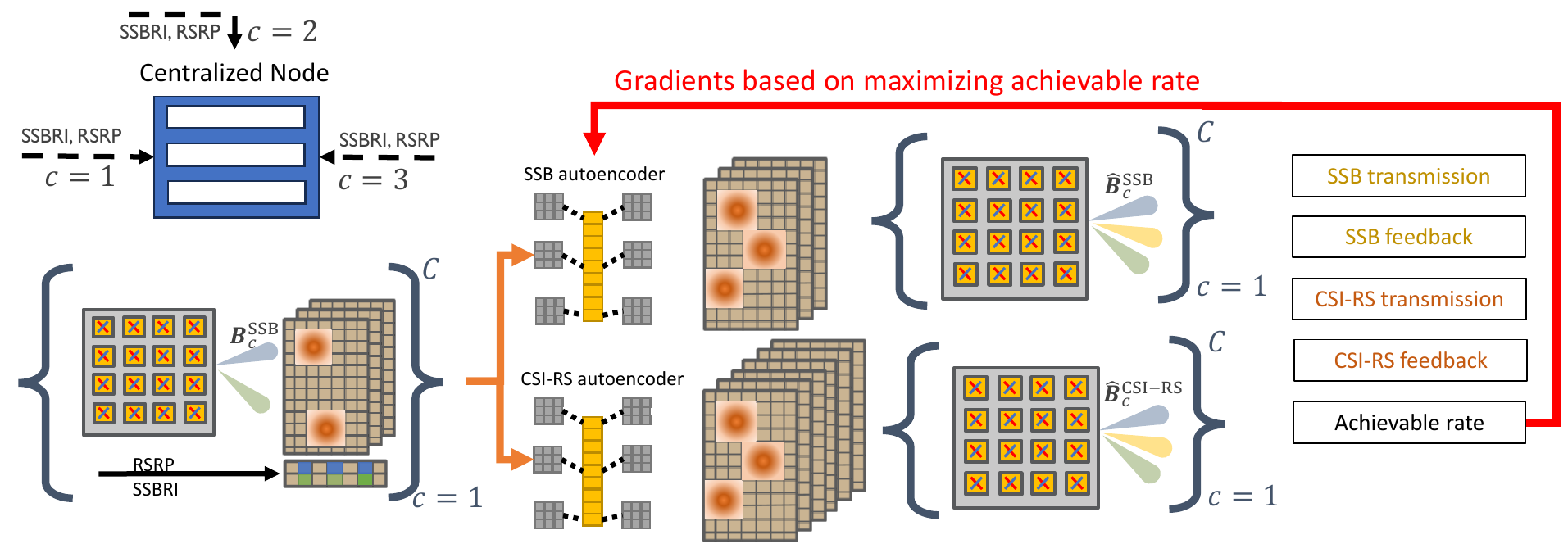}
        \caption{A visual representation of the $\algo$ inference and training within the network. SSB feedback is shared between cells to jointly generate SSB and CSI-RS codebooks. Gradients are backpropagated through the entire beam management system to maximize the achievable rate during CSI-RS.}
        \label{fig: system}
    \end{figure*}

    In this paper, we propose a gradient-complete multi-cell MU-MIMO codebook optimization strategy. We previously found that neural codebook design based on beamspace representations is a powerful tool for codebook design \cite{Dreifuerst2023CodebookFeedback}. Training the architecture to maximize beam training power was also effective \cite{Dreifuerst2023hierarchical}, but ultimately focused on an intermediate goal--beam training power--rather than the network goal of spectral efficiency maximization. Conceptually, in this paper, we reframe the learning objective to maximize the sum spectral efficiency in the presence of multi-user and multi-cell interference. In this section, we define the end-to-end neural architecture as shown in Figure \ref{fig: system}, beginning with the beamspace preprocessing and including up through codebook generation. We define three important aspects of the neural architecture design process that fundamentally aid the $\algo$ algorithm. 
    
    \textbf{Design aspect 1:} \emph{Encapsulating the beam management framework within the end-to-end learning produces $\algo$ codebooks that maximize system-level performance.}
    Learning SE maximizing codebooks requires more than simply training the neural network to predict a specific set of beamformers in a typical supervised fashion. Instead, we capture the beam management process as described in Section \ref{sec: system} in the backpropagation step. Then we train the neural network to maximize the achievable rate during CSI-RS based on \eqref{eqn: ach_rate}. Ultimately, because of the reliance on the CSI-RS codebook for analog beamforming in the data transmission, the codebooks used during beam training should support high data rates rather than simply rank-$1$ transmissions.

    The first step in the neural processing pipeline is converting the previous SSB Codebook into a beamspace representation \cite{Dreifuerst2023CodebookFeedback, Dreifuerst2023hierarchical}. The beamspace representation provides a consistent learning paradigm where the SSB feedback corresponds to an ``image'' of the beamforming magnitude over a range of azimuth and elevation directions. In addition to setting up the learning paradigm, the beamspace representation also allows the neural architecture to generalize to any antenna size or number of users by separating those dimensions from the inputs and outputs of the neural network. To be precise, neural networks are often limited by requiring consistent dimensions for inputs and outputs but we circumvent this through the beamspace representation which allows for arbitrarily sizing the input ``image'' to sampling dimensions $\Nxo\times\Nyo$ in the azimuth and elevation directions for each beamformer. The only requirement is that $\Nxo\ge\Nx$ and $\Nyo\ge\Ny$ to ensure the beamspace is not aliased. The beamspace transformation can be performed at each cell and shared with local cells or a small centralized node during each SSB period ($\approx20$ms) to form a \textit{weakly-centralized} setting (see Figure \ref{fig: system}). The centralization or SSB sharing is motivated by the long SSB periods as well as only sharing the SSB feedback between cells, rather than large packets of information. 
    
    We define the beamspace representation for each cell $c$ as $\Obsc_{c}\in\complex^{(\Nxo+1)\times(\Nyo+1)\times(\Lmax)}$ formulated in the same way as \cite{Dreifuerst2023hierarchical}. First, transformation matrices $\vect{U}_{N_{1}, N_{2}}$ are calculated from the array responses $\vect{a}_{N}(\theta)$ \cite{HeathOverviewSP4mmWave2016} for antenna dimension $N_{1}$ to beamspace sampling dimension $N_2$ 
    \begin{align}
        \boldsymbol{\theta}_{N_2} &= \frac{1}{\pi}[0, 1, ..., N_2-1]^T\\
        \vect{U}_{N_{1}, N_{2}} &\triangleq [\vect{a}_{N_{1}}(\boldsymbol{\theta}_0), ... \vect{a}_{N_{1}}(\boldsymbol{\theta}_{N_{2}-1})] \label{eqn: ang_matrix}.
    \end{align}
    Then the beamspace representation is
    \begin{align}
        \widehat{\Obsc}_{c, i} &= \vect{U}_{\Nxa, \Nxo}^* \Fssb_{c,i} \vect{U}_{\Nya, \Nyo} \quad \forall c, i.  \label{eqn: beamspace}\\
        \vect{O}^{\text{BSC}}_c &= \left\{\left[\widehat{\Obsc}_{c, i}, \sum \vect{1}_{\vect{m}=i}, \vect{1}_{\vect{m}=i}^T\vect{p}\right]\right\}_{i=0}^{\Lmax-1}. \label{eqn: lastOBSC}
    \end{align}
    Note that in a centralized scenario, the codebook generation is carried out at a shared compute node and shared with each cell which in turn share the SSB feedback as shown in Figure \ref{fig: system}. In the event of a fully disaggregated scenario, the interfering cells cannot be jointly optimized, the nearby cell's beamspace representations are not shared, and interference instead is treated as noise during the gradient updates.

    \textbf{Design aspect 2:} \emph{Beamspace representations enable generalization in dynamic codebook learning algorirthms.}
    The beamspace preprocessing forms a consistent basis that abstracts the carrier frequency and antenna geometry from the neural network. The abstraction is performed by the beamspace conversion that incorporates the array response, which is a function of the operating frequency and geometry. The beamspace representation allows for the same neural network to be used in a range of deployment strategies for network operators \cite{Dreifuerst2023hierarchical}. An evaluation of the generalization capabilities is presented in Section \ref{sec: Sim_results}.
    
    After computing the beamspaces for each cell, the concatenated $\Obsc$ is fed into a $3+3$ layer convolutional autoencoder architecture with ($128, 128, 256$) filters and $2\times2$ max pooling between layers. In contrast to our prior work \cite{Dreifuerst2023hierarchical}, the final $3$ layers perform an inverse convolution (typically termed convolution-transpose in deep learning frameworks) instead of using fully connected layers due to the large dimensional outputs which would require significant hardware resources (more than a single GPU) to parameterize in a fully connected architecture. By employing parameter sharing during the $3$ encoding and $3$ decoding layers, the network only requires $\approx 2$ million parameters for the joint SSB and CSI-RS prediction for all cells. 

    The output of the neural network is a predicted beamspace which is then converted back to the complex codebook entries for the next beam training period. The deconversion process is performed in the same way as the conversion process using the matrix inverses of \eqref{eqn: beamspace}. With the new codebooks $\Fssbpred,\Fcsipred$ we can introduce the end-to-end learning strategy for rate-maximized codebook learning. Ultimately, the end-to-end strategy allows for learning complex and arbitrary codebooks that minimize the loss in achievable rate from a perfect CSI, SU-MIMO, interference-free setting. 

    First, the target ``labels'', which are the maximum achievable spectral efficiencies, are computed from the channel assuming perfect CSI and SU-MIMO data transmission. The maximum achievable SE is calculated from the singular value decomposition (SVD) for a user $u$ with connected BS $c$ as
    \begin{align}
        \vect{U}_{u}, \vect{S}_{u}, \vect{V}^*_{u} &= \vect{H}_{c, u, t, k}\\
        r_{u} &= \sum_{\nr=1}^{\Nr}\log_2\left(1 + \frac{S_{u, \nr}^2}{\sigma^2}\right) \label{eqn: svd}.
    \end{align}
    For ease of notation, we focus on a single time-frequency resource $t, k$, although the actual implementation is performed over an applicable set of resources. In a single-cell scenario, the system could then calculate the achievable spectral efficiency resulting from serving all active users with different analog precoder selections.
    When extending this to the multi-cell setting, however, it is pessimistic to assume that every analog precoder $\Bactive$ from every BS is active and causing interference. Furthermore, directly assuming all precoders cause interference ignores the timing aspects of beam management, where interference is acceptable during beams that a user does not end up selecting. In essence, the beams from nearby cells should cover non-overlapping regions at a given time. We capture this property by assuming the interference only occurs from the beams used by the interfering BS during the CSI-RS of the selected CSI-RS $\ihat$ as in \eqref{eqn: SINR}. Mathematically the achievable spectral efficiency can therefore be calculated as \eqref{eqn: ach_rate}.

    \textbf{Design aspect 3:} \emph{Incorporating realistic time-frequency usage within the training allows the algorithm to coordinate beamforming between interfering cells.} 
    Interference during CSI-RS causes channel estimation errors and CSI-RS beams may be allocated by users for the nearby cells. Assuming interference only during the user-selected CSI-RS resource frees the algorithm to learn analog precoders that do not interfere strongly during the same CSI-RS to improve channel estimation and subsequent data transmission. In a non-centralized format, the neural network would be limited to simply maximizing the signal strength, but this does not necessarily improve the network ESSE (see Figure \ref{fig: ML_ESSE} of Section \ref{sec: Sim_results}). Then the overall loss function is the mean-squared-error (MSE) of the target and predicted achievable rates
    \begin{align}
        \mathcal{L}(\vect{r}, \widehat{\vect{r}}) = \frac{1}{U}\sum_u (r_u - \widehat{r}_u)^2. \label{eqn: e2e_loss}
    \end{align}
    In an ideal scenario, the loss would approach $0$ with perfect CSI and no noise. Of course, there is always non-zero noise and interference thereby causing the loss function to be strictly positive. Gradient calculation is backpropagated throughout the entire system model for the selected SSB and CSI-RS codewords. 
    

\section{Raytraced beam management dataset}
\begin{figure*}[!t]
        \centering
        \subfloat[Scene 1 \label{fig: scene1}]{\includegraphics[width=3.25in]{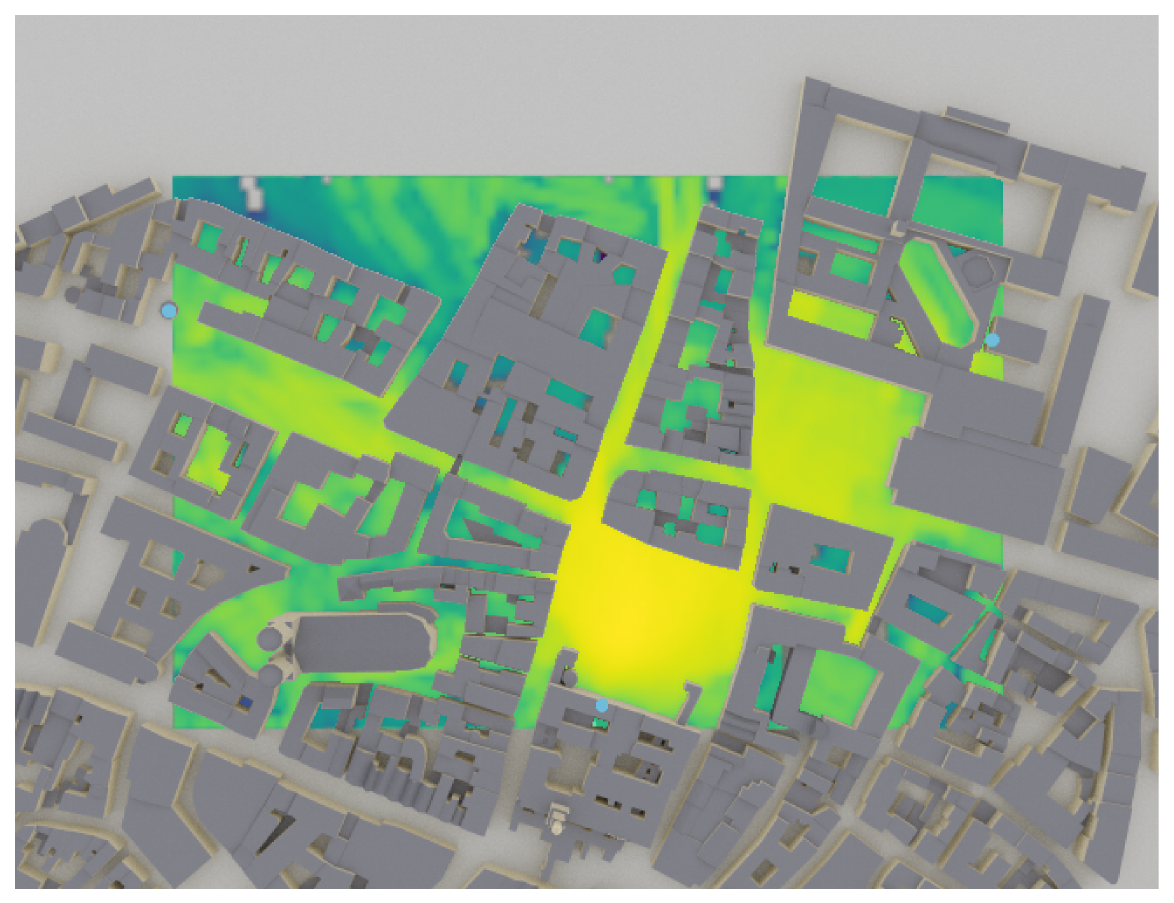}}
        \subfloat[Scene 2 \label{fig: scene2}]{\includegraphics[width=3.25in]{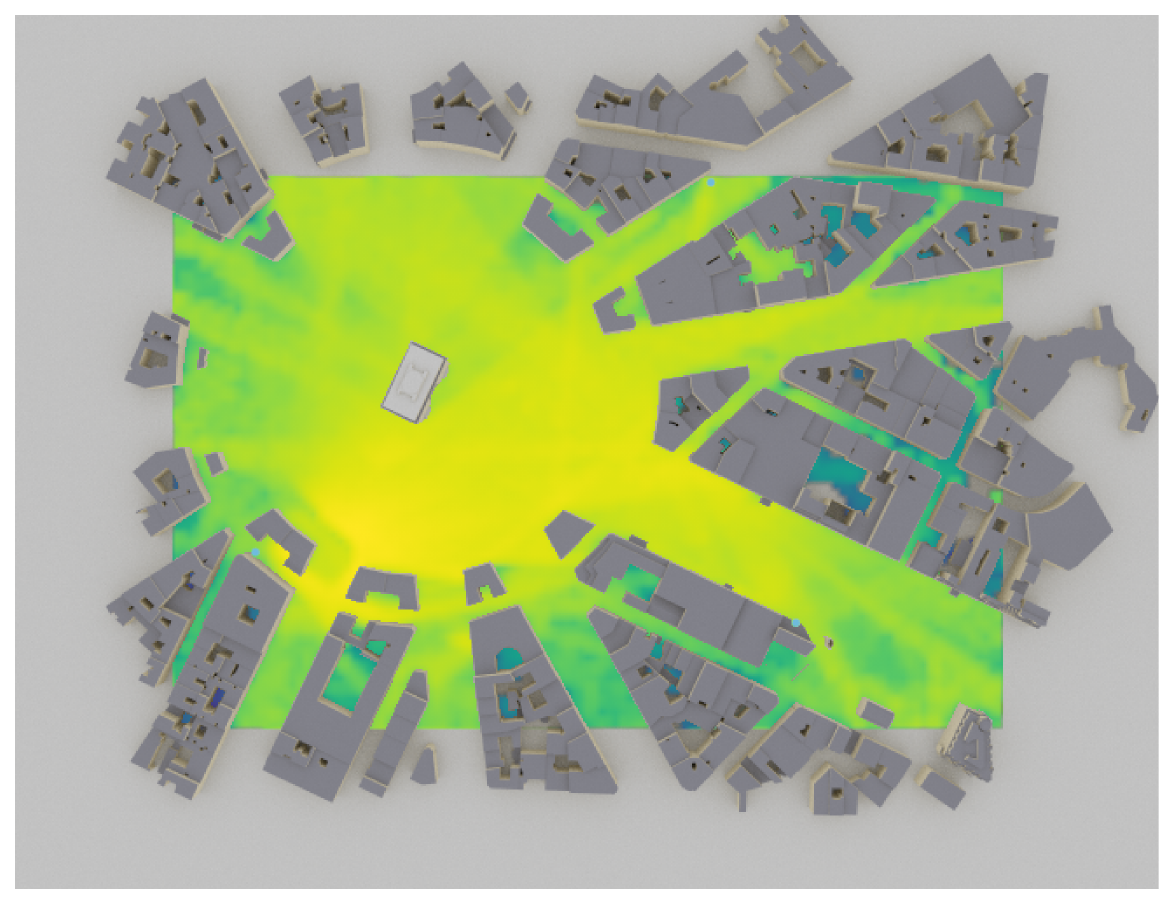}}
        \caption{A comparison of the powermaps in two environments with different base station heights and carrier frequencies. Scene 1 corresponds to Munich, Germany operating at $10$GHz and base stations mounted at $40$m tall and $450$m intersite distance. The OOD environment in scene 2 is Paris, France operating at $20$ GHz, with $27$m high serving cells and $370$m intersite distance. A low resolution power map is rendered on each scene in the overlapping region from the perspective of base station $1$ in each of the scenes.}  
        \label{fig: scenes}
    \end{figure*}

    System-level simulations that capture the entire beam management cycle are critical for evaluating the ultimate network effects of the proposed $\algo$ codebooks. Simulation can be broken down into two steps: channel generation and system modeling. We employ Sionna \cite{sionna}, a raytracing channel simulator, that generates an offline dataset of user channels to sample from. In our results, we define a default scenario of a raytraced environment around the Frauenkirche in Munich, Germany operating at $10$ GHz with three base stations $40$m tall each equipped with a planar array $\Nx=16$, $\Ny=16$, and $32$ digital ports facing toward the serving region. We also consider a scenario with a smaller antenna array, $\Nx=8$, $\Ny=8$, and another possible deployment scenario with the system operating in Paris, France at $20$ GHz from a height of $27$m. These scenarios, shown in Figures \ref{fig: scene1}-\ref{fig: scene2}, allow us to consider urban environments in the low- and high-bands of the proposed upper-mid band \cite{Holma2021XMIMO} as well as different array geometries and performance in the face of OOD data. Users are assumed to have $\Nr=4$ antennas and scattered outdoors throughout the scene at heights of $0.5$-$1.5$m above ground. The channels are sampled every millisecond over a $100$MHz bandwidth comprised of $270$ resource blocks with $30$kHz subcarrier spacing. 

    After generating channels for $1000$ mobile UEs, a training dataset of beamspaces, channels, and target spectral efficiencies can be built up. For each dataset sample, a random number of active users is selected $U \sim \mathcal{U}[8, 20]$ and drawn from the channel dataset. Of those users, each is assigned a $20\%$ probability of being considered a ``new'' user, and feedback of these users is not included in the subsequent beamspace calculation. Non-new users are associated with base stations based on the SSB feedback $\vect{b}$ determined from the prior SSB codebook, which is initialized with a DFT codebook. The beamspace is also generated from the prior SSB codebook as well as the per-cell SSB feedback. The SEs are calculated according to \eqref{eqn: svd} and the triplet (beamspace, channels, and SE) are stored for the offline training. During training, the new SSB and CSI-RS codebooks are generated from the beamspace, SSB and CSI-RS feedback is simulated for the next beam training period, and the models are updated based on gradients calculated from the end-to-end loss \eqref{eqn: e2e_loss}. Gradient updates are performed with an Adam optimization and learning rate of $10^{-4}$ from a training dataset with a validation dataset used for determining and comparing model performance. All simulation results are calculated from a new, unseen test dataset. Overall, the network is trained with $20,000$ samples in the training datasets corresponding to different array sizes, environments, and carrier frequencies. 

    During model evaluation, new channels are generated, the AI model is used to predict the codebooks, and the data transmission performance is calculated according to a greedy scheduler with RZF precoding \eqref{eqn: F_RF}-\eqref{eqn: ESSE}. In particular, the base stations estimate the achievable sum SE for every combination of users from the coarse CSI reported and select the highest combination, i.e. there is no cooperation outside of the codebook generation. The lack of coordination results in a straightforward scheduling algorithm using only the available, imperfect CSI. The ESSE \eqref{eqn: ESSE} is then calculated from the true CSI for evaluating the overall performance of the proposed system. The resulting RSRP, SINR, and ESSE metrics are used for comparison in Section \ref{sec: Sim_results}. DFT codebooks are used for comparison where the SSB codebook and CSI-RS codebook are wide and narrow DFT codebooks. In the DFT codebooks, the least commonly used beams are not considered. For example, beams angled upward are never used as these would never be selected by users scattered at ground level. The DFT selection provides us with a reasonable comparison with traditional codebooks while still incorporating a small degree of site-specific codebooks.

\section{Simulation results} \label{sec: Sim_results}
    In this section, we present a careful evaluation of the proposed $\algo$ algorithm. Neural networks are capable of function matching in a wide range of tasks, but also susceptible to overfitting \cite{Vuckovic2023GeneralBeamSel, ZhangDLinWirelessSurvey2019}. We consider scenarios where the training data and testing data come from different distributions as a result of different physical array geometries, deployment locations, and operating frequencies. The first set of results focuses on a site-specific scenario to evaluate the best-case performance; out-of-distribution settings are tested in the later subsections. Unless specified, we assume $\Lmax=16$, $\Ncsi=16$, type-II PMI codebooks with $\Lcsi=4$ \cite{Dreifuerst2023magazine, Qin2023CSIFeedback}, modulation and coding scheme $0$ is used during uplink feedback transmission, and all results refer to Scene $1$ from Figure \ref{fig: scenes}.
    
                
    \subsection{SSB RSRP}
\begin{figure}[!t]
            \centering
            \includegraphics[width=3.25in]{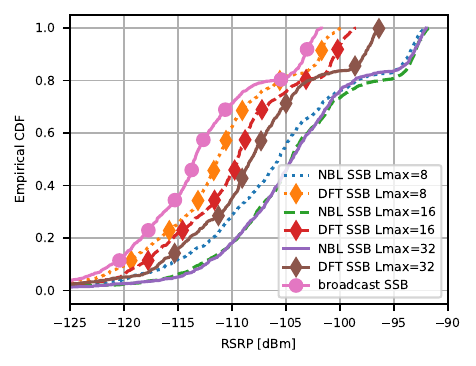}
            \caption{The CDF of the reported RSRP using $\algo$ and DFT codebooks of different sizes $\Lmax$. The broadcast SSB refers to a single, non-beamformed signal and is shown for comparison. The $\algo$ codebooks result in an average of $>5$dB gain over DFT codebooks and reduced user outage rates.}
                \label{fig: SSB_RSRP}
        \end{figure}

        The first evaluation focuses on the RSRP performance of the proposed $\algo$ algorithm. The RSRP is evaluated using the SSB codebooks. There are two goals: 1) To achieve a sufficiently strong connection so that UEs are capable of accurate synchronization and 2) That the subsequent CSI-RS subset $\Bactive$ achieves a beamformed channel with high SINR for channel estimation and MU-MIMO SE. The first goal is directly quantified by the reported RSRP, with the goal typically being that all users are above the sensitivity, which is conservatively $\approx-120$dBm. We show the empirical CDF of the reported RSRP in Figure \ref{fig: SSB_RSRP} with different numbers of SSB beams $\Lmax$. It can be seen that the DFT codebook produces twice as many users below the sensitivity and more than $5$dB separates the bottom $20\%$ of users. We can also see that $\algo$ codebooks achieve better performance with the smallest codebooks $\Lmax=8$ than DFT codebooks ever reach as a result of the rich scattering environment. Current 5G implementations are limited to $\Lmax=8$ for sub-6GHz bands and $\Lmax=64$ in mmWave bands, so ML codebooks can help alleviate beam training overhead with X-MIMO arrays in 6G. 

\begin{figure}[!t]
            \centering
            \includegraphics[width=3.25in]{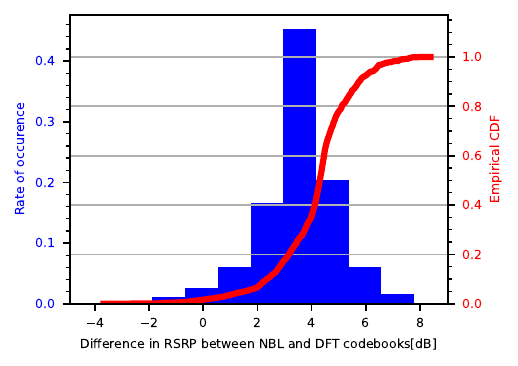}
            \caption{A comparison of the reported RSRP by users with $\algo$ codebooks and DFT codebooks. Less than $2\%$ of users reported a decreased RSRP with $\algo$ codebooks, while more than $90\%$ of users gain more than $3$dB.}
            \label{fig: SSB_delta}
        \end{figure}

        We also show a comparison of the per-user difference in RSRP between the two codebook methods in Figure \ref{fig: SSB_delta}. While the ECDF shows large-scale differences, it is also important to characterize if there are negative effects of the $\algo$ codebooks. There are significant improvements as a result of the $\algo$ codebooks with $93\%$ of users increasing by more than $3$dB. Yet, a small amount ($\le 2\%$) of users experience a worse RSRP compared to DFT codebooks. Given the slight deficit among a small percentage of the population, this further validates the improvements in the SSB codebook performance as a result of learned codebooks. We also compare the load balancing of codebooks in Table \ref{tb: BS_alloc}. Overall there is not a significant difference in user allocation because it is primarily a function of the environment, where Cell $2$ experiences NLOS channels for most user locations (see Figure \ref{fig: scene1}). 
        
                    
\begin{table}[!t]
\centering
\caption{User allocation by cell ID}
\begin{tabular}{|c|c|c|c|}
\hline
\textbf{Codebook} & \textbf{\begin{tabular}[c]{@{}c@{}}Cell 0 User\\ Allocation\end{tabular}} & \textbf{\begin{tabular}[c]{@{}c@{}}Cell 1 User\\ Allocation\end{tabular}} & \textbf{\begin{tabular}[c]{@{}c@{}}Cell 2 User\\ Allocation\end{tabular}} \\ \hline
NBL                & 46.6\%                                                                      & 45.8\%                                                                      & 7.5\%                                                                       \\ \hline
DFT                & 49.8\%                                                                      & 42.2\%                                                                      & 8\%                                                                         \\ \hline
\end{tabular} \label{tb: BS_alloc}
\end{table}

    \subsection{CSI-RS performance}
\begin{figure}[!t]
            \centering
            \includegraphics[width=3.25in]{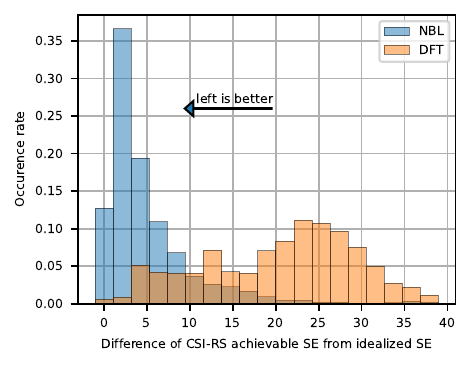}
            \caption{A histogram of the difference in achievable and ideal SE. The ideal SE is achieved with perfect CSI and a fully digital array. The difference shows how much SE is lost with respect to the achievable rate during CSI-RS from \eqref{eqn: ach_rate}. The difference in effective SINR translates to more than $2/3$ of users receiving CSI-RS within $5$dB of the ideal achievable SE.}
            \label{fig: SE_training}
        \end{figure}

        While the $\algo$ SSB codebooks noticeably improve the initial access results, evaluating the second goal of supporting CSI-RS with high SE requires a joint evaluation with the CSI-RS. In contrast to initial access, CSI-RS are allocated or designated based on the users already in the network. As such, the candidate CSI-RS codebook should improve the link configuration and support high data rate communications. In the first CSI-RS results, we compare the difference between perfect CSI SU-MIMO SE and the achievable SE from \eqref{eqn: ach_rate}. The performance difference is equivalent to the training loss function \eqref{eqn: e2e_loss} and captures the performance and timing relationship as a result of the codebook design process. The results, depicted in Fig. \ref{fig: SE_training} highlight the value of AI/ML codebooks over traditional codebooks for beam management, pushing the achievable SE to within $10$bps/Hz of the idealized SE in over $80\%$ of results. 
        
\begin{figure}[!t]
            \centering
            \includegraphics[width=3.25in]{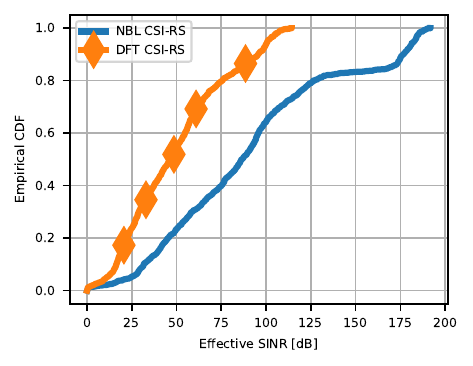}
            \caption{A statistical CDF comparison of the effective SINR obtained during CSI-RS reception.  It can be seen that the $\algo$ codebooks show a gain of more than $40$dB for the majority of users.}
            \label{fig: CSIRS_SINR}
        \end{figure}

        While the achievable SE is a critical metric, it is valuable to understand where the SE gains are coming from. The performance gains are especially important if a new system is considered with different algorithms, for example with a different channel estimation technique. Multi-cell interference during CSI-RS causes errors in channel estimation--and subsequent data transmission--so mitigating interference during CSI-RS is important for maximizing performance. We introduce an additional metric, the effective SINR, to help characterize the CSI-RS performance. The effective SINR is determined as the equivalent SINR needed to obtain the given SE in dB as
        \begin{align}
            \text{Eff-}\SINR_{u} = 10 \log_{10}(2^{\text{SE}^{\text{CSI-RS}_{\hat{i}_u}}_u}-1).
        \end{align}
        The effective SINR, which is shown in Figure \ref{fig: CSIRS_SINR}, also allows for breaking down the contributions to the achievable rate due to improved signal combining and interference-noise-ratio.
        
\begin{figure}[!t]
            \centering
            \includegraphics[width=3.15in]{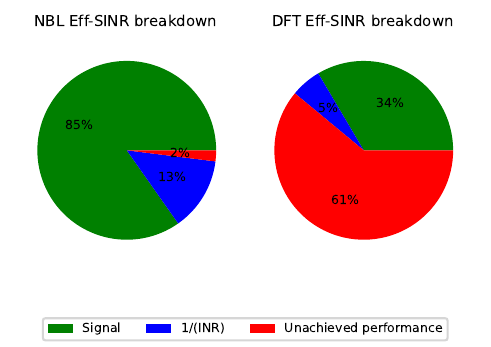}
            \caption{A ratio of the average effective signal power and inverse-interference-noise (1/INR) power compared to the achievable effective SINR. The effective SINR difference between $\algo$ and DFT codebooks is significant, but it can be seen that this is a result of both improving signal quality and reducing interference, up to nearly the optimal SINR.}
            \label{fig: CSIRS_S2I}
        \end{figure}

        The proposed $\algo$ shows a substantial improvement in effective SINR compared to DFT codebooks. The performance uplift is from the two-fold problem: improving the beamforming coherent combining and reducing the interference from nearby cells during overlapping transmission periods. In fact, if we highlight the contributions between signal strength and inverse interference, the results become even more clear as shown by Figure \ref{fig: CSIRS_S2I}. In this case, the effective signal power is significantly improved, although the interference is also cut in half. 
       
    \subsection{Network characterization} \label{sec: model_update}
\begin{figure}[!t]
            \centering
            \includegraphics[width=3.25in]{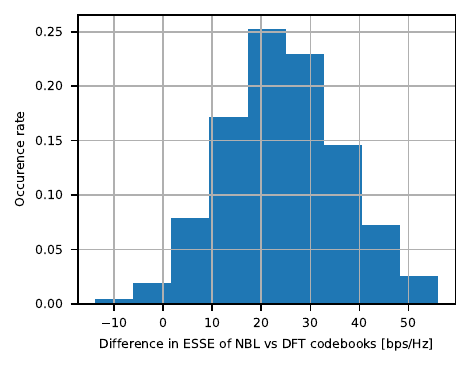}
                \caption{Visualization of the difference in ESSE between the proposed $\algo$ and DFT codebooks in the network. Almost universally $\algo$ codebooks show a significant gain in the achievable rate. In comparison, less than $3\%$ of simulations show any decrease in performance.}
            \label{fig: ESSE}
        \end{figure}

        High achievable SE during CSI-RS, however, does not necessarily imply the wideband, MU-MIMO data transmission will reflect a similar improvement. In the final set of evaluations, we seek to characterize the proposed $\algo$ performance as seen by a network operator. All of the prior results focused on achievable SE as a result of the beamformed channel--i.e. did not include channel estimation error, quantization error, resources outside of the CSI-RS, or MU-MIMO interference. Although DFT codebooks significantly underperformed during CSI-RS, the beamformed channel with DFT codebooks is more likely to be spatially sparse and see less performance loss from the quantization and limited feedback than learned codebooks. In this subsection, we focus on the ESSE following the entire beam management process. First, the ESSE gain as a result of $\algo$ codebooks in the site-specific case is shown in Figure \ref{fig: ESSE}. From these results, it is clear there is a valuable performance gain using $\algo$ codebooks, especially considering the ESSE is then multiplied by the channel bandwidth of $100$MHz to get the total network rate difference, which is over $2.5$Gbps improvement on average. Further, less than $3\%$ of simulations show a loss of performance using the proposed codebook algorithm.

\begin{figure}[!t]
            \centering
            \includegraphics[width=3.25in]{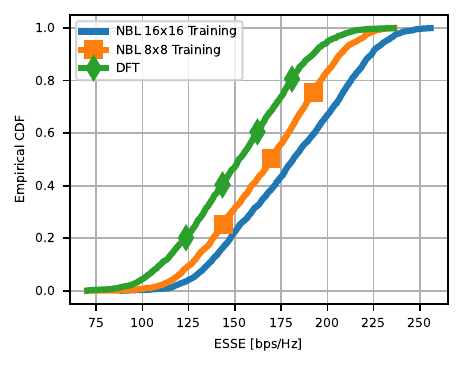}
                \caption{Statistically comparing the network ESSE using different training data geometries. In all cases, the codebooks are tested with data corresponding to $\Nx=16, \Ny=16$ arrays. The benefit of beamspace learning is a natural generalization across geometries allowing for training and testing for different array sizes or even arbitrary configurations \cite{Dreifuerst2023hierarchical}. Even with the challenge of ``upscaling'' the array function, the beamspace translation enables significant gains over traditional codebooks during network upgrading.}
            \label{fig: geometry}
        \end{figure}

        Next, we consider a typical network operator upgrade scenario where data is available from a 5G antenna array (for example $\Nt=64$) that is used to train the model and test with a 6G X-MIMO array ($\Nt=256)$ in Figure \ref{fig: geometry}. Because of the beamspace transformation, the same neural network can be applied for different arrays which is a valuable benefit for both operators upgrading or potential OpenRAN settings. It can be seen that there is a $\approx10$bps/Hz performance loss as a result of the array upscaling, but that the performance is still highly effective compared to generic codebooks.

        While array geometry is an interesting generalization, an even more common scenario is the generalization across scenes. We propose a comparison where the training and testing raytraced environments are different, and the operating carrier frequency is shifted from $10$GHz to $20$GHz. Inherently, the beamspace transformation should extend well across frequencies due to the integration of the beamspace transformation with the frequency-dependent array response, however, the frequency change also impacts how the signals interact with the environment. OOD formulations are known to be challenging for AI/ML in wireless \cite{Vuckovic2023GeneralBeamSel} but even in OOD channels, the network ESSE distribution shows a substantial benefit as a result of $\algo$ codebooks. These results validate both the beamspace representation as a generalizable preprocessing strategy as well as the overall learned model's capabilities.

        \begin{figure}[!t]
            \centering
            \includegraphics[width=3.15in]{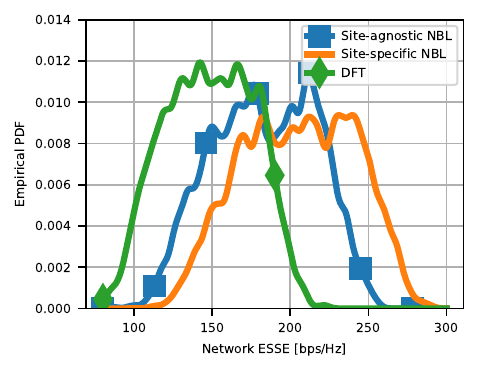}
                \caption{The empirical PDF of the network ESSE for $\algo$ and DFT codebooks. The site-agnostic $\algo$ is trained on data from the Munich scene and tested in the Paris scene. The site-specific $\algo$ is trained and tested on data from Paris. It can be seen that $\algo$ generalizes to the new scene to outperform DFT codebooks, although there is still a performance loss.}
            \label{fig: agnostic_ESSE}
        \end{figure}

\begin{figure}[!t]
        \centering
        \includegraphics[width=3.15in]{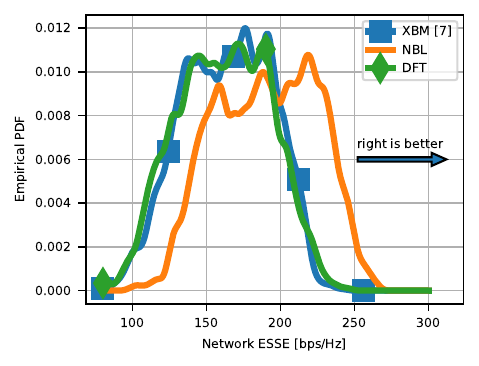}
            \caption{A comparison of the XBM model from \cite{Dreifuerst2023hierarchical} with our proposed solution. The XBM model is given the same training data although each cell operates and learns an independent XBM realization. Our results align with those from \cite{Dreifuerst2023hierarchical} where the beam alignment gains do not translate to improved ESSE, especially with inter-cell interference.}
        \label{fig: ML_ESSE}
    \end{figure}

    In the final results, we compare the $\algo$ with prior work \cite{Dreifuerst2023hierarchical}, which was based on learning single-cell power-maximizing codebooks without considering interference. We had previously found that power-maximizing codebooks showed modest gains over DFT codebooks in ESSE, but ultimately the gains were lower than expected based on the beam training performance. We show the relative performance of the two models in Figure \ref{fig: ML_ESSE}, where the X-BM algorithm from \cite{Dreifuerst2023hierarchical} is trained independently for each of the three cells. Note that the parameter sizes are not comparable--the prior X-BM algorithm requires over $100$ times more parameters to employ the proposed architecture. Yet, even with a much smaller network, the change in optimization goal and slight coordination results in a valuable performance improvement. We see that the new $\algo$ formulation overcomes the lack of system performance of the XBM model. 
    All told, the proposed algorithm outperforms traditional codebooks throughout the beam management process and generalizes across a wide range of use cases and deployment strategies.


\section{Conclusion}
    In this paper, we consider a weakly-centralized AI/ML algorithm for beam management in X-MIMO systems. The proposed strategy learns SSB and CSI-RS codebooks for overlapping serving cell regions from beamspace representations. By integrating neural networks into the codebook design process, the cellular network can reduce overhead and obtain stronger wireless links than generic codebooks. Furthermore, codebook design can be integrated with interference-aware metrics to improve overall network performance with hybrid arrays. 

    The proposed $\algo$ algorithm incorporates an end-to-end learning strategy where codebooks are evaluated and gradients are computed based on the difference with the per-user optimal SE. In combination, codebook learning relies on beamspace transformations to convert arbitrary beamforming vectors into images representing the magnitude of the array response in a given direction. By integrating this domain knowledge, the $\algo$ can efficiently learn codebooks that generalize across array geometries, environments, and operating bands. We show that the learned codebooks improve network ESSE by more than $25\%$ over traditional codebooks.

    The results of this investigation highlight important results with respect to ML/AI for beam management and 6G integration. First, we found that integrating AI/ML into the codebook design process can reduce the need for larger codebooks with X-MIMO arrays. Secondly, we found that although learned beams include arbitrary beamforming directions, the interference is actually reduced compared to DFT codebooks. This combination of reduced beam training overhead, reduced interference, and improved beam alignment all while generalizing effectively showcases the potential benefits of using AI/ML for beam management. In future work, we will consider further realistic systems subject to hardware impairments, timing and frequency offsets, and performance impacts as a result of other system design choices.

\bibliographystyle{IEEEtran}
\bibliography{IEEEabrv, RMD_all_refs, heath_refs_all}
\end{document}